\documentclass[a4paper,usenatbib,twocolumn]{mnras}

\pdfoutput=1
\usepackage[T1]{fontenc}
\usepackage{ae,aecompl}

\usepackage{graphicx}	
\usepackage{amsmath}	
\usepackage{amssymb}	

\newcommand{\beq}[0]{\begin{equation}}
\newcommand{\eeq}[0]{\end{equation}}

\title[Jets shining through their stripes]
{GRB and blazar jets shining through their stripes}

\author[D.~Giannios \& D.~Uzdensky]{
D.~Giannios,$^{1}$\thanks{E-mail: dgiannio@purdue.edu}
D.~A.~Uzdensky,$^{2}$\thanks{E-mail: uzdensky@colorado.edu}
\\
$^{1}$Department of Physics, Purdue University, 525 Northwestern Avenue, West Lafayette, IN 47907, USA\\
$^{2}$Center for Integrated Plasma Studies, Physics Department, 
   390 UCB, University of Colorado, Boulder, CO 80309, USA
}

\date{\bf Draft version: \today }

\pubyear{2018}

\begin{document}
\label{firstpage}
\pagerange{\pageref{firstpage}--\pageref{lastpage}}
\maketitle

\begin{abstract}
%
%
Black-hole driven relativistic astrophysical jets, such as blazars and gamma-ray bursts (GRBs), are powerful sources of electromagnetic radiation. Their emission is powered by some energy dissipation and particle acceleration mechanism operating in-situ at large distances from the black hole.  We propose that the formation of the dissipative structures in the jet is controlled by the time variability of the accretion disc. We argue that the  open magnetic field lines through the black hole, which drive a strongly magnetized jet, may have their polarity reversing over time-scales related to the growth of the magneto-rotational dynamo in the disc. Consequently, the jet is characterized by an alternating toroidal field polarity along its propagation axis, i.e., it is a "striped jet". Magnetic reconnection in the current sheets that form between the stripes dissipates the alternating-field energy and thus powers further jet acceleration. Here, we consider a jet with a broad distribution of stripe widths $l>l_{\rm min}$, above a dominant scale~$l_{\rm min}$. We find that the bulk acceleration of the jet, driven by the annihilation of the stripes, is very gradual. The dissipation rate peaks at a distance $z_{\rm peak}\sim 10^6 R_g\, (\Gamma_\infty/30)^2\, (l_{\rm min}/1000R_g)$, where $R_{g}$ is the black-hole's gravititional radius and $\Gamma_{\infty}$ the {jet's} asymptotic Lorentz factor, and exhibits a very broad plateau extending by $\sim 4-5$ orders of magnitude in distance. The prolonged energy dissipation accounts for the flat-to-inverted long-wavelength spectra commonly observed in jets. The model can also account for the broad range of flaring timescales of blazars and the fact that their bulk acceleration appears to continue out to $\sim 100$~pc scales. In GRB jets, the model predicts comparable power for the photopsheric and Thomson-thin emission components.

\end{abstract}






\section{Introduction}
\label{sec-intro}

Jets from black holes are ubiquitous. 
They shine to us across all wavelengths from radio to gamma-rays from several different classes of sources, such as stellar-mass black holes in galactic X-ray Binaries (XRBs), supermassive black holes in active galactic nuclei (AGN), and perhaps core-collapse and neutron-star merger gamma-ray bursts (GRBs). Particularly spectacular are blazar jets --- powerful ultrarelativistic (with typical bulk Lorentz factors $\Gamma\sim 30$) AGN jets pointing directly at us. Relativistic abberation effects in these jets increase their apparent luminosity and compress their observed variability timescales, making them appear extremely bright and rapidly variable. The same relativistic effects, but even stronger, act in GRB jets ($\Gamma \sim 300$). In both cases, however, precisely because the jets are oriented along our line of sight, we see them as unresolved, point-like gamma-ray sources. This prevents us from using direct imaging to tell where --- i.e., how far from the central engine --- the dominant gamma-ray emission is produced. The current estimates for the location of this so-called {\it "blazar zone"} (in the case of blazars) or the "prompt gamma-ray emission zone" (in GRBs; hereafter we use the term "jet emission zone" for both classes of sources) are therefore highly uncertain and based only on indirect arguments.  For example, for blazars they range from $\sim 0.01$~pc to $\sim 30$~pc (Dermer \& Schlickeiser 1994; Sikora, Begelman \& Rees 1994; Georganopoulos \& Kazanas 2003; Ghisellini \& Tavecchio 2008; Agudo et al. 2011). Developing a solid understanding of the location of the dominant gamma-ray emission in these relativistic jets is an important challenge to both theory and observation.

Furthermore, since the ultrarelativistic electrons (and possibly positrons) responsible for emitting the gamma rays are highly energetic, their radiative cooling time is often much shorter than the travel time from the central engine. This means that the jet emission zone has to be powered by some {\it in-situ} energy dissipation and particle acceleration mechanism, and thus the location of the jet emission points to the main site of {\it energy dissipation} in relativistic jets.

More generally, however, one is interested not just in the location of  the main site of energy dissipation but in the entire {\it distribution of energy dissipation} along the length of the jet. Observationally, this is manifested as the distribution of bolometric radiation power along the jet and can in principle be probed, e.g., by observing non-blazar AGN jets that do not point directly along our line of sight and are thus spatially resolved. This includes many classical FR-I jets, one of the most important examples of which is the jet in~M87 (Biretta et al. 1991; Marshall et al. 2002). AGN jets and microquasars are characterized by flat-to-inverted long-wavelength emission, with the resolved emission region being more compact at higher frequencies. This emission pattern is understood as synchrotron emission from a conical jet in which the characteristic self-absorption frequency decreases with distance along the jet axis (Blandford \& K{\"o}nigl 1979; hereafter BK79). The phenomenologically very successful BK79 model implies continuous particle acceleration over a very broad range of length scales along the jet axis. 
In addition, another, complimentary approach to probing the axial profile of energy dissipation in jets, including in blazars and GRBs, is to use the rich {\it timing information} that is available in different spectral bands. For instance, blazar jets are strongly variable, with prominent flares in various spectral bands that last over very different time-scales, ranging from as long as years to as short as several minutes. The broad range of flaring time-scales indicates that, instead of having a single scale, the dissipation process in jets is gradual, taking place over a range of distances from the black hole.

The present paper aims to address these important questions from the theoretical perspective, by analyzing the processes responsible for the formation of dissipative structures in the jet. In particular, we tie the location of these structures with the distance that the plasma in the jet travels over the time it takes to form and dissipate them. In our model, the time of injection of these structures is in turn controlled by the time variability (e.g., the power spectrum) of the central engine, e.g., a black-hole accretion disc. The accretion process itself is believed to be driven by the magneto-rotational instability (MRI) operating in the disc (Balbus and Hawley 1991). In this picture, the MRI-driven dynamo generates loops of magnetic field of varying polarity over time-scales comparable to the orbital time in the inner parts of the disc. The accreting gas advects the inner footpoints of the magnetic loops into the black hole while the outer footpoints still thread the disc. The resulting differential rotation of the two footpoints forces the loops to open up rapidly (Uzdensky 2005, Parfrey et al. 2015, see also Barkov \& Baushev 2011). A rotating black hole threaded by open magnetic field lines drives a strongly magnetized jet (Blandford \& Znajek 1977). 
The polarity of the loops that accrete subsequently is, in general, varying. As a result, the polarity of the  open magnetic field through the black hole reverses (or fluctuates) over a range of time-scales related to the  characteristic MRI growth time in the disc, which in turn scales with the Keplerian period. Consequently, the jet is characterized by varying toroidal field polarity along its propagation axis. 
We refer to this structure as a "striped jet" due to its similarity to a striped wind that is formed from the ballerina-skirt equatorial current sheet in oblique pulsar magnetospheres (Coroniti~1990). 

The dynamics of a striped jet has been studied by Drenkhahn 2002 (hereafter D02) and Drenkahn \& Spruit (2002) in the context of GRB jets under the assumption of a single unique stripe width dominating the jet (as expected, for instance in the oblique magnetar model for GRBs (Usov 1992; Uzdensky \& MacFadyen 2007; Metzger et al. 2011). They argued that, while the jet expands, there is plenty of time for the stripes to dissipate the free energy of the reversing magnetic field through { the process of} magnetic reconnection. The energy released by reconnection drives both the jet bulk acceleration and particle energization. When the expansion time-scale of the jet equals that of the magnetic reconnection over the whole stripe width, the dissipation process is complete and the jet bulk speed saturates. In our case, however, we expect the stripe widths to follow a broad distribution determined by the temporal power spectrum of MRI-driven dynamo at different scales in the disc. One can then anticipate that the distribution of widths of the magnetic stripes will be reflected in the axial dissipation profile in the jet. We therefore generalize the~D02 and Drenkhahn \& Spruit (2002) analysis to this situation.
  
This paper is organized as follows. 
In Section~\ref{sec-picture} we describe the basic idea of a striped jet and make simple estimates for the location of the jet emission zone, which we find to consistent with those in the literature. We also briefly discuss the distribution for stripe widths fed into the jet from the central engine (Section~\ref{subsec-plateau}). In Section~\ref{sec-analysis} we analyze the distribution of energy dissipation and the bulk acceleration profile along the jet  and present an analytical solution of this problem. We demonstrate that dissipation remains rather flat over a very broad (several decades) range of scales roughly centered logarithmically about the main dissipation-zone distance and that the bulk acceleration saturates at the outer edge of the dissipation zone. In Section~\ref{sec-astro} we discuss the two most important astrophysical applications of our model:  to blazar jets (Section~\ref{subsec-astro-blazars}) and to GRBs (Section~\ref{subsec-astro-GRBs}). Section~\ref{sec-disc} is devoted to a broader discussion. Finally, in Section~\ref{sec-conclusions} we draw our conclusions and outline the prospects for future research.


\section{\bf Basic Physical Picture}
\label{sec-picture}

Blazar or GRB emission is powered at some distance from the central-engine black hole,  determined by the plasma physics of energy dissipation and particle acceleration. We postulate that dissipation is magnetic in nature, in the form of magnetic reconnection (see, e.g., Spruit, Daigne \& Drenkhahn 2001; Giannios et al. 2009). We aim to investigate where along the jet it takes place. 

Magnetic reconnection requires formation of current sheets. 
In principle there are several main pathways to developing current sheets in astrophysical jets (e.g., McKinney \& Uzdensky 2012): 

(1) Magnetic field irregularities/structures produced at the base of the jet near the central engine and then simply advected into the jet while strengthening and steepening along the way. This may lead to the formation of a striped {jet}.

(2) Nonlinear development of current-driven instabilities, e.g., the kink instability (Eichler 1993; Begelman 1998; Giannios \& Spruit 2006).

These two scenarios are, of course, not mutually exclusive and both may happen in the same system. 

In this paper we focus on the first scenario.
In particular, we envision a striped jet with magnetic field reversals that are produced at its base by the black-hole accretion disc.  Indeed, recent magnetohydrodynamic (MHD) simulations of accretion-disc turbulence driven by the MRI show that MRI dynamo in stratified discs produces quasi-periodic reversals of the magnetic flux that is emerging vertically out from the disk into the wind and jet (Davis, Stone \& Pessah 2010, O'Neill et al. 2011, Simon, Beckwith \& Armitage 2012).  This large-scale dynamo cycle manifests itself through a butterfly diagram (in $zt$ coordinates) for the emerging magnetic flux, and is conceptually similar to the {familiar} 11-year solar-dynamo cycle that also produces large-scale field reversals. 

According to the simulations, the characteristic period for these reversals, $\tau_{\rm rev}$, is about 10 orbital periods (cf., in the solar dynamo case, the reversal period is longer, about 200 solar rotation periods). We expect most of the energy powering the jet to come from the inner part of the disc, in the range of radii comparable to the innermost stable circular orbit (ISCO) $R_{\rm ISCO}$, i.e., several gravitational radii~$R_g \equiv GM/c^2$. 
Thus, for rough estimate, we take $R_{\rm base}  \simeq {\rm a \ few} \times R_g \,$. 
The corresponding {Keplerian} orbital period at these radii is 
$T_{\rm base} = 2\pi \Omega_K^{-1} (R_{\rm base}) = 2\pi (R_g/c) (R_{\rm base}/R_g)^{3/2}$.
Taking, e.g.,  $R_{\rm base} = 10 R_g$, 
we get $T_{\rm base} \simeq 200 R_g/c$, and hence $\tau_{\rm rev} \sim 10^3 R_g/c$.
Thus, the jet will have quasi-periodic toroidal field reversals of axial ($z$) length (in {the} lab frame; see also \S~2.1 for more discussion) of about 
\beq
l_{\rm rev} \simeq c \tau_{\rm rev}\sim 10^3 R_g \,\label{eq-lrev} .
\eeq 
Note that for rapidly spinning black holes, possibly launching the most powerful jets, the ISCO is closer to~$R_g$. For these sources, $l_{\rm rev} \sim 10^2 R_g$ may be a more accurate estimate. 

Let us now ask what should happen to these field reversals as they propagate with the jet to large distances.  At first, they are long and tall columns (of length $l_{\rm rev}$ and radius $\sim R_{\rm base} \ll l_{\rm rev}$)  of reversing toroidal field, stacked on top of each other. 
However, the jet is not cylindrical. It expands sideways as it propagates outward. 
Its shape,  $R_j(z)$,  may in general be a complicated function, governed, e.g., by the external pressure profile (see, e.g., Lyubarsky 2010). The external pressure can serve to both collimate the jet and accelerate it to ultrarelativistic speeds. Here for simplicity we will consider  a jet that, after its initial collimation and acceleration phases, becomes conical with a fixed opening angle $\theta_j$ (typically, of order a few degrees, roughly 0.1~rad).  The cylindrical radius of the jet  then increases as $R_j(z) = \theta_j z$. Importantly, this radius becomes larger than the width $l_{\rm rev}$ of the stripes at a distance 
\beq
z =\theta_j^{-1} l_{\rm rev} \sim 10^4 R_g \, .
\label{eq-z-pancake}
\eeq  
Beyond this distance, instead of tall and slender cylindrical segments of alternating toroidal field, we are dealing with a stack of thin flying pancakes. As a result of their large  aspect ratio, the stripes are becoming prone to tearing instabilities and fast magnetic reconnection.

We shall now proceed with our estimate of how long it takes {the stripes of alternating toroidal field} to annihilate each other by reconnection. Since {the flow in the jet is ultrarelativistic}, we will do this by first going to the jet's comoving frame, estimating the reconnection time there, and then going back to the laboratory frame. We will denote the quantities in the jet frame with primed symbols. 

As an illustration, in this section we will for simplicity assume a fixed constant jet bulk Lorentz factor~$\Gamma_j$, of order $\sim$~10--30, as observed in AGN jets, {or $\sim 300$, typical of GRB jets}. (In the next section, we will relax the constant-$\Gamma_j$ assumption by consistently treating the jet acceleration  powered by magnetic dissipation.)  Then, the width (in the $z$-direction) of stripes in the comoving frame can be estimated as 
\beq
l'_{\rm rev} = \Gamma_j l_{\rm rev} \, .
\eeq

Assuming that fast collisionless reconnection quickly ensues and destroys adjacent stripes, the reconnection time in this frame is 
\beq
t'_{\rm rec}  = l'_{\rm rev} / v_{\rm rec} = l'_{\rm rev}/\beta_{\rm rec}c = \beta_{\rm rec}^{-1} \Gamma_j  l_{\rm rev}/c \, , \label{eq-trec}
\eeq
where $\beta_{\rm rec} \equiv v_{\rm rec}/c  = E_{\rm rec}/B_0 \simeq 0.1$ is the dimensionless relativistic reconnection rate (in the next Section we provide a more general parameterization of the reconnection rate depending of the plasma magnetization).

Now, going back to the laboratory frame, we find that relativistic time dilation results in 
\beq
t_{\rm rec} = \Gamma_j t'_{\rm rec} = \beta_{\rm rec}^{-1}\Gamma_j^2 l_{\rm rev}/c  \, .
\eeq
{This gives $t_{\rm rec} \sim 10^3\, l_{\rm rev}/c$ for blazars and $t_{\rm rec} \sim 10^6\, l_{\rm rev}/c$ for~GRBs.} Similar arguments have been presented by Lyubarsky \&  Kirk (2001) for pulsar striped wind and D02 and Drenhahn \& Spruit (2002) for gamma-ray burst (GRB) jets. Thus, the  lab-frame length of the { jet} dissipation zone of primary magnetic irregularities associated with field reversals injected into the jet at the base -- which we associate with the size and location of the blazar zone -- is 
\beq
z_{\rm rec} = c t_{\rm rec}  \simeq \beta_{\rm rec}^{-1} \Gamma_j^2  l_{\rm rev} 
\sim 10^3 \beta_{\rm rec}^{-1} \Gamma_j^2  R_g   
\sim 10^4 \Gamma_j^2  R_g \, ,
\eeq
where we assumed $\beta_{\rm rec} = 0.1$ for a rough estimate. As a consistency check, we see that, for ultrarelativistic jets, $\Gamma_j\gg 1$,  the dissipation zone $z_{\rm rec}$ is bigger than the scale given by equation~(\ref{eq-z-pancake}).

Applying this estimate for AGN/blazar jets with $\Gamma_j \sim 10$, we get 
\beq
z_{\rm rec}^{\rm AGN} \sim 10^6  R_g  \, .
\eeq
Thus, for example, for a typical blazar-powering black hole of $M = 10^8 M_\odot$, corresponding to $R_g \simeq 1.5 \times 10^8\, {\rm km} = 1.5 \times 10^{13}\, {\rm cm} = 1\, {\rm AU}$, we get $l_{\rm rev} \simeq 1000\, {\rm AU}$, and 
$z_{\rm rec}^{\rm AGN} \sim 10^6\,  {\rm AU} \simeq 5 \, {\rm pc}$. 

{In the case of a typical GRB jet with} $\Gamma_j \sim 300$, and $M = 10 M_\odot$, we get 
\beq 
z_{\rm rec}^{\rm GRB} \sim 10^9  R_g \sim 10^{15}{\rm cm} \, .
\eeq
Both of the above estimates are quite compatible with existing {observational and theoretical} constraints on the blazar zone and GRB emission radius, respectively (see Section~\ref{sec-astro}).

\subsection{A dissipation plateau over a broad range of scales}
\label{subsec-plateau}

The above arguments provide an estimate for a characteristic distance from the black hole at which the dissipation rate peaks. In fact, as we show qualitatively in this and quantitatively in the next Section, in the striped jet model the dissipation remains substantial over a very broad range of scales, both smaller and larger than~$z_{\rm rec}$.

At smaller scales $z<z_{\rm rec}$ the dissipation is only partial. Since magnetic dissipation and bulk acceleration are closely connected (D02; see also next Section), 
the fraction $f_{\rm rec}$ of the total Poynting flux that has been dissipated at a distance $z$ also provides an estimate for the bulk Lorentz factor of the jet as a fraction of its asymptotic value:  $\Gamma \simeq f_{\rm rec} \Gamma_{\infty}$, where $f_{\rm rec}\simeq t'_{\rm exp}/t'_{\rm rec}$ and $t'_{\rm exp}\sim z/\Gamma c$ is the comoving expansion time-scale of the jet and $\Gamma_\infty$ is the jet's terminal Lorentz factor after complete conversion of Poynting flux to the kinetic energy flux.
Using equation~\ref{eq-trec} for $t'_{\rm rec}$, we find that $\Gamma\sim (\Gamma_{\infty}z\beta_{rec}/l_{\rm rev})^{1/3}$, which is, within an order-unity numerical factor, the same as the classic result of D02 and Drenkhahn \& Spruit (2002).  One can verify that $\Gamma\simeq\Gamma_{\infty}$ for $z \simeq z_{\rm rec}$. The acceleration is very gradual, $\Gamma \propto z^{1/3}$, and the same is true for the $z$-profile of the dissipation rate~(D02). The exact solutions derived in the next section show an even flatter dissipation profile for $z<z_{\rm rec}$. 

Our estimate (equation~\ref{eq-lrev}) for $l_{\rm rev}\sim 10^3 R_{\rm g}$ corresponds to the shortest width $l_{\rm min}$ of the stripes, generated by the MRI dynamo in the inner part of the accretion disc. In a more realistic picture, however, MRI turbulence is active over a broad range of radii in the disc, thus introducing magnetic field reversals over a broad range of time-scales. To study the effect of the various time-scales injected into the jet, we introduce the concept of the normalized distribution of field-reversal time-scales. 
The Fourier transform of this quantity corresponds to the power-spectrum of the poloidal magnetic flux at the jet base. For definiteness, in this paper we will assume that this distribution is a truncated power law: 
\beq
\mathcal{P}(\tau)\equiv \frac{{\rm d}P}{{\rm d}\tau}= \frac{1}{(a-1)\tau_{\rm min}}\left(\frac{\tau}{\tau_{\rm min}}\right)^{-a}, \quad \tau\ge \tau_{\rm min},
\eeq
with a constant power-law index~$a>1$. 
The index $a$ could in principle be determined by global 3D disk simulations that {could} properly resolve the MRI dynamo in the disk as well as in the accretion disk corona (Uzdensky \& Goodman 2008). Our rough guess for $a$ is $a=5/3$. This comes from associating the time-scales generated at distance $R$ with the local orbital period $\tau \propto T \propto R^{3/2}$ and ascribing the power in proportion to the gravitational energy {released} locally, $\mathcal{P} \propto 1/R$. Therefore d$P$/d$\tau$=(d$P$/d$R$)(d$R$/d$\tau$)$\propto \tau^{-5/3}$.

\begin{figure}
\begin{center}
\includegraphics[width=0.5\textwidth]{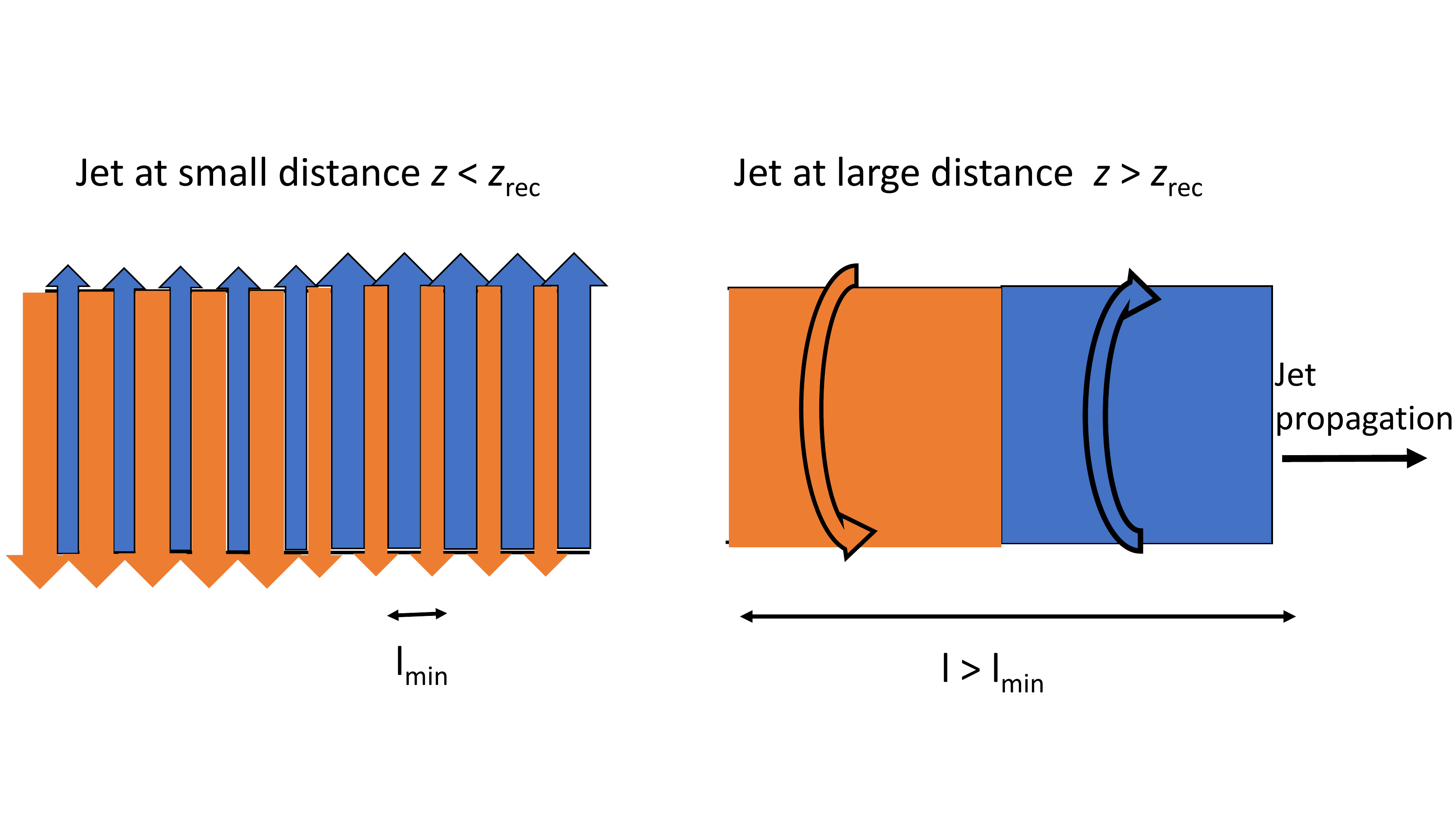}
\end{center}
\caption{Cartoon of striped jet with two stripe width scales $l$, $l_{\rm min}$, where $l\gg l_{\rm min}$. The jet (for simplicity depicted as cylindrical) propagates horizontally towards the right direction. At small distance from the black hole $z<z_{\rm rec}$ (left), it contains stripes of both widths $l_{{\rm min}}$, $l$. Orange (blue) color denotes the regions with downward (upward) polarity for the magnetic field. Note that the widths of the two regions are not equal. At large distance from the black hole  $z>z_{\rm rec}$ (right) only the long stripes survive.  
\label{fig-cartoon}
}
\end{figure}

As a result of the time-scale distribution of magnetic field reversals~$\mathcal{P}(\tau)$ at the jet base, the jet is characterized by a corresponding distribution of stripe widths along its propagation axis,  $\mathcal{P}(l)$.  To visualize this point, consider the case where the polarity of the magnetic field threading the black hole changes periodically over a well defined  time~$\tau$. As a result, at any fixed distance $z$ from the black hole the magnetic field in the jet reverses polarity with a frequency $1/\tau$. Since the striped structure travels with the jet, the stripe width in the lab frame is $v(z)\tau$ where $v(z)$ is the speed of the jet.  As long as the jet has already accelerated to a speed close to the speed of light, the relation between stripe with $l$ and injection time $\tau$ is, therefore, simply $l=c\tau$. We thus assume that the jet contains a distribution of stripe widths that extends to $l \gg l_{\rm min}=c\tau_{\rm min}$ (see Fig.~\ref{fig-cartoon}), following the same distribution as that describing the disc time-scales:
\beq
\mathcal{P}(l)\equiv \frac{{\rm d}P}{{\rm d}l}= \frac{1}{(a-1)l_{\rm min}}\left(\frac{l}{l_{\rm min}}\right)^{-a}, \quad l\ge l_{\rm min}.\label{eq-width}
\eeq


As we have argued, given a width for the field reversal of $l_{\rm rev}=l_{\rm min}$, there is a corresponding distance along the jet $z_{\rm rec}\simeq \beta_{\rm rec}^{-1} \Gamma_j^2  l_{\rm min} $ at which this stripe is dissipated (see Section~\ref{subsec-model-dynamics}). In a jet that contains stripes over a range of widths $l>l_{\rm min}$, dissipation proceeds to distances $z>z_{\rm rec}$ for these wider stripes (see Fig.~\ref{fig-cartoon}). The energy dissipation {$z$-profile} then closely tracks the stripe distribution: $z$d$L_{\rm diss}/{\rm d}z\propto z^{1-a}$, i.e., declines gradually at {large scales} (see next section for a more accurate calculation of the dissipation profile).


\section{Analysis}
\label{sec-analysis}

\subsection{Dynamics of acceleration}
\label{subsec-model-dynamics}

The problem of a relativistic striped jet with a single unique field-reversal scale was discussed by Drenkhahn in~D02. Here we briefly rehearse his approach and generalize it to a jet with a distribution of stripe widths, and provide an analytic solution.

To make the model analytically tractable, we postulate: 
\begin{itemize}
\item {\it The flow is conical, and the magnetic field is predominantly toroidal.} 
The jet is injected at some initial distance $z_{\rm o}$ with an initial bulk Lorentz factor $\Gamma_{\rm o}\ll \Gamma_{\infty}$. 
The flow is assumed cold at the injection distance, i.e.,  the internal (thermal) energy density of the plasma in the jet at {this point} can be neglected compared to both the magnetic and rest-mass energy densities. The jet is, therefore, injected magnetically dominated, with Poynting-to-kinetic energy flux ratio $\sigma_{\rm o}=\Gamma_{\infty}/\Gamma_{\rm o}-1\gg 1$ (see also equation~\ref{eq-sigma-u} below). The steady state assumption also means that jet magnetization and the total (kinetic + magnetic) power remain constant in time at the injection distance.

Our injection setup implies that we do not consider the initial acceleration and collimation stages of the jet close to the black hole. Implicitly, we assume the following, rather standard, physical picture. The jet undergoes magneto-centrifugal acceleration and collimation through interactions with  the surrounding gas. It is accelerated to the fast-magnetosonic speed and beyond during these early stages (Tchekhovskoy, McKinney \& Narayan 2008; Komissarov et al. 2009; Lyubarsky 2010). When the ambient gas density or pressure profile become sufficiently steep, the jet turns conical and, after a possible episode of abrupt "rarefaction acceleration", ideal-MHD acceleration essentially freezes (Tchekhovskoy et al. 2010; Komissarov et al. 2010). Our focus {here}  in this article is on the jet dynamics at this later stage. While both ideal and dissipative MHD mechanisms can in principle operate at similar scales in a more realistic model, here we ignore this possibility.

\item {\it The jet contains stripes of reversing polarity in the toroidal magnetic field.} 
The widths $l$ of the stripes (measured in the lab frame) follow a truncated power-law distribution~$\mathcal{P}(l)\propto l^{-a}$, see eq.~(\ref{eq-width}). As we argued in the previous section, the stripe-width distribution along the jet~$\mathcal{P}(l)$ closely tracks the time-scale distribution~$\mathcal{P}(\tau)$ of polarity reversals at the jet base. As long as the jet has accelerated to a speed close to the speed of light, the relation between stripe {width} $l$ and injection time $\tau$ is simply $l=c\tau$.

\item {\it The stripes are separated by current sheets where the magnetic energy dissipates by reconnection} with a characteritic reconnection inflow speed (in the jet comoving frame) $v_{\rm rec} = c \beta_{\rm rec}$, assumed to be a fixed fraction $\epsilon$ of the comoving Alfv\'en speed~$v_A$, to be defined below. { Numerous studies of collisionless magnetic reconnection indicate that $\epsilon \simeq 0.1$ and so we adopt this value for quantitative estimates presented in this paper. As long as the jet has not approached its terminal Lorentz factor $\Gamma_\infty$, it remains  Poynting-flux-dominated, i.e., $v_A\simeq c$; in this case, one deals with relativistic reconnection, characterized by $\beta_{\rm rec} \simeq \epsilon \simeq 0.1$.}
For arbitrary magnetization, we follow D02, parameterizing the reconnection rate as $\beta_{\rm rec}=\epsilon v_A/c$.

\item {\it The flow remains cold throughout the jet even beyond the injection point, despite dissipation. This assumption means that most of the dissipated energy is used to drive bulk acceleration.}

In the limit of a steady, conical relativistic flow with toroidal field of a fixed polarity, the jet coasts at a constant velocity (Michel~1969). The presence of magnetic polarity reversals enables dissipation of magnetic energy. The {resulting} steepening magnetic pressure profile in the jet drives its bulk acceleration (Lyubarsky \& Kirk 2001; D02; Kirk \& Skj{\ae}raasen 2003).
In reality, only about one half of the released magnetic energy is converted to heat, i.e., to the internal plasma energy, with the other half  being directly converted to bulk flow kinetic energy (see Spruit \& Drenkhahn 2004). However, the thermal/internal energy  also gets quickly converted to bulk kinetic energy anyway because the gas pressure drives the expansion of the flow and therefore  becomes diminished by adiabatic expansion. The cold flow assumption effectively just skips this intermediate step of conversion of the internal energy to bulk kinetic energy.

The above argument for thermal-to-kinetic energy conversion neglects radiative losses in the flow. Drenkhahn \& Spruit (2002) relaxed the cold flow assumption and allow for radiative losses. They showed that, when the radiative losses are small, the thermal energy of the plasma in the jet always remains a modest fraction (at most $\sim 1/4$) of the bulk kinetic energy. If, on the other hand, all the energy injected {into} particles is radiated away rapidly (i.e., on a time-scale much shorter than the expansion time of the jet), approximately half of the jet power is radiated away, somewhat reducing the terminal jet speed (see also Levinson \& Globus 2016 for the effect of Compton drag on the dynamics and radiative efficiency of the jet and Zrake \& Arons 2017 for the effect of turbulence in the flow). For all practical purposes, the flow can be assumed to remain approximately cold throughout  even when radiative losses do not significantly affect the dissipation and acceleration dynamics of the jet. 

\end{itemize}

Under these assumptions, the jet flow is characterized by two conserved (i.e, independent of~$z$) quantities: the total electro-mechanical power of the jet $L$ and the mass flux~$\dot{M}$. 
The total jet power has two components, kinetic and Poynting: $L=L_k+L_p=L_k(1+\sigma)=\Gamma \dot{M}c^2(1+\sigma)$. In the last expression $\sigma=L_p/L_k$ is the jet magnetization. 
The magnetization also determines the comoving Alfv\'en 4-velocity $u_A\equiv \Gamma_A\beta_A \simeq \sqrt{\sigma}$. 
The luminosity-to-mass-flux ratio $L/\dot{M}c^2$ is a measure of the maximum bulk Lorentz factor that the jet can achieve if all the magnetic energy is converted into kinetic: $\Gamma_{\infty}=L/\dot{M}c^2$.
The last expression can also be written as
\beq
\Gamma_{\infty}=\Gamma(z)\, [1+\sigma(z)],
\eeq
where one can see that, while the dissipation proceeds reducing the jet magnetization, the bulk $\Gamma$ increases and asymptotically approaches $\Gamma_\infty$ for $\sigma\to 0$. The magnetization $\sigma$ can then be conveniently expressed as a function of the 4-velocity~$u$ 
\beq
\sigma(z) = \frac{\Gamma_\infty}{\Gamma} - 1\simeq \frac{u_\infty}{u(z)} - 1 \,
\label{eq-sigma-u}
\eeq
where in the last step we used the approximation $u\simeq \Gamma$ which is valid in the ultrarelativistic regime of interest here.

The Poynting luminosity converted to the flow's kinetic luminosity per unit length along the jet at a given distance~$z$ equals to
\beq 
\frac{{\rm d}L_{\rm diss}}{{\rm d}z}=\frac{{\rm d}L_{k}}{{\rm d}z}=\frac{{\rm d}\Gamma}{{\rm d}z}\dot{M}c^2
\eeq 
{\it Therefore, in this model, magnetic dissipation and bulk acceleration are closely related -- with the former driving the latter. The last expression quantifies their relation.}

Assuming a single width $l_{\rm rev}$ for all the stripes, D02 worked out the differential equation that describes the jet 4-velocity $u=\Gamma \beta \approx \Gamma$ as a function of distance~$z$:
\beq 
\frac{{\rm d}u}{{\rm d}z}=\frac{2\epsilon}{l_{\rm rev}}\frac{(u_\infty-u)^{3/2}}{u^2u_{\infty}^{1/2}}.
\label{eq-D02}
\eeq 
Note that the $(u_\infty-u)^{3/2}$ factor implies that the acceleration ceases when $u\to u_\infty$. Also, in the limit $l_{\rm rev}\to \infty$, i.e., uniform polarity jet, there is no acceleration of the flow (Michel 1969).

Now we proceed to generalize this expression to account for a distribution of stripe widths $\mathcal{P}(l)$. The processes of bulk acceleration and dissipation are closely connected, with the dissipation been complete when $u\to u_\infty$. When the jet has reached a given 4-velocity~$u$, a fraction $(u_\infty -u)/u_\infty$ of its luminosity remains in the form of Poynting flux.  Only stripes with widths $l>l_{\rm eff}$ such that 
\beq 
\int_{l_{\rm {eff}}}^{\infty}\mathcal{P}(l){\rm d}l =\frac{u_\infty -u}{u_\infty} \quad {\rm or} \quad \frac{{\rm d}u}{{\rm d}l_{\rm eff}}=  u_\infty \mathcal{P}(l_{\rm eff})
\label{eq-pl-int}
\eeq 
survive in the jet.  We expect that the reversals with the shortest width are the first to dissipate. As the jet propagates out and accelerates, the characteristic stripe width $l_{\rm eff}$ increases. Given a choice for the distribution $\mathcal{P}(l)$, equation~(\ref{eq-pl-int}) can be integrated and solved for~$l_{\rm eff}(u)$. The differential equation that describes the jet acceleration is given by equation~(\ref{eq-D02}) but with  $l_{\rm rev}$ replaced by $l_{\rm eff}(u)$:
\beq 
\frac{{\rm d}u}{{\rm d}z}=\frac{2\epsilon}{l_{\rm eff}(u)}\frac{(u_\infty-u)^{3/2}}{u^2u_{\infty}^{1/2}}.
\label{eq-accel-diff}
\eeq 

Equations (\ref{eq-pl-int})-(\ref{eq-accel-diff}) describe the jet acceleration for arbitrary distribution $\mathcal{P}(l)$. Analytical treatment of the problem is possible for simple forms of~$\mathcal{P}(l)$. In particular, we consider a range of stripe widths that follow a power-law distribution: $\mathcal{P}(l)\propto l^{-a}$, for $l>l_{\rm min}$. Equation~(\ref{eq-pl-int}) can then be analytically integrated and solved for $l_{\rm eff}(u)=l_{\rm min}(1-u/u_\infty)^{1/(1-a)}$.
Using equation~(\ref{eq-accel-diff}), the jet acceleration is then described by:
\beq 
\frac{{\rm d}u}{{\rm d}z}=\frac{2\epsilon}{l_{\rm min}}\frac{(u_\infty-u)^{\frac{1-3a}{2-2a}}}{u^2u_{\infty}^{1/2}}.
\label{eq-dudz_a}
\eeq 
In the limit of $a\to \infty$, and setting $l_{\rm min}=l_{\rm rev}$, the last expression is equivalent to the D02 case (eq.~\ref{eq-D02}).

Equation~(\ref{eq-dudz_a}) can be written in a more compact dimensionless form by defining $\chi \equiv u/u_\infty$, $k\equiv (1-3a)/(2-2a)$ and $\zeta\equiv 2z/z_{\rm rec}=2\epsilon z/l_{\rm min}u_\infty^2$:
\beq 
\frac{{\rm d}\chi}{{\rm d}\zeta}=\frac{(1-\chi)^k}{\chi^2}.
\label{eq-chi_zeta}
\eeq 
This differential equation can be solved analytically by separation of variables, yielding an implicit expression for~$\chi(\zeta)$:
\beq 
\frac{(1-\chi)^{3-k}}{k-3}-\frac{2(1-\chi)^{2-k}}{k-2}+\frac{(1-\chi)^{1-k}}{k-1}=
\zeta+ C \, , 
\label{eq-chi_zeta-impl}
\eeq 
where $C$ is the integration constant.
This expression assumes that $k\ne 1, 2, 3$ (in which cases logarithmic terms arise). 
The integration constant $C$ is determined by the conditions at the inner injection boundary $\zeta=\zeta_{\rm o}$. For simplicity, in the rest  of this section we set $\chi_{\rm o}=\zeta_{\rm o}=0$, yielding  the constant $C=(k-3)^{-1}-2(k-2)^{-1}+(k-1)^{-1}$. We note however that our results do not depend sensitively on this choice and in Section~\ref{sec-astro} we will use more realistic injection conditions as expected for blazar and GRB jets.

For the interesting specific case $a=5/3$ (a value motivated by our argument stated in \S~2.1), the corresponding $k=3$. The acceleration solution $\chi(\zeta)$ is then given implicitly by $-2(1-\chi)^{-1}+0.5(1-\chi)^{-2}-\ln(1-\chi)=\zeta-1.5$; it can be easily solved numerically. As we show in the following, the properties of the solutions vary smoothly with varying~$a$. As a fiducial reference value, we adopt $a=5/3$ but also investigate the dependence of our results on the $a$ parameter.

An explicit solution for the acceleration profile is possible for $a=1.4$ which corresponds to $k=4$. For this choice, equation~(\ref{eq-chi_zeta-impl}) reads: $(\zeta+1/3)(1-\chi)^3-(1-\chi)^2+(1-\chi)-1/3=0$. This cubic equation for $\chi$ has a simple solution 
\beq 
\chi(\zeta)=\frac{3\zeta-(3\zeta)^{2/3}+(3\zeta)^{1/3}}{1+3\zeta}\quad {\rm for}\quad k=4.
\label{eq-chi_zeta-k4}
\eeq 


\begin{figure}
\begin{center}
\includegraphics[width=0.5\textwidth]{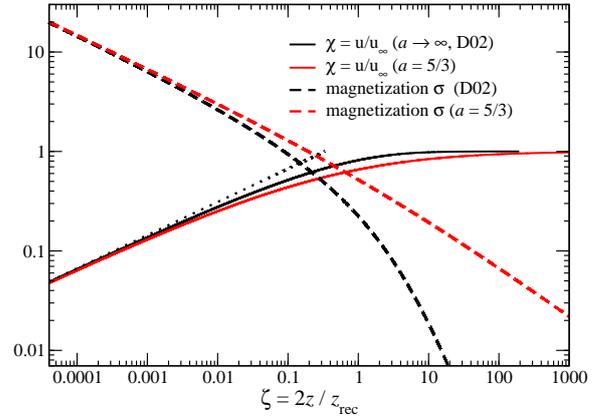}
\end{center}
\caption{Jet bulk 4-velocity $\chi = u/u_\infty$ (solid lines) and magnetization $\sigma$ (dashed lines) as a function of distance along the jet. {The black and red curves correspond to the cases of single-width stripes  ($a\rightarrow \infty$, as in D02) and a stripe-width distribution with power-law index $a=5/3$, respectively.}
The black dotted line shows the asymptotic solution $\chi \sim \zeta^{1/3}$ (eq.~\ref{eq-u_z-small-z}) valid a $\zeta\ll 1$. 
The smaller the $a$ value, the more gradual the acceleration of the jet.
\label{fig-gamma}
}
\end{figure}


The general solution for the jet acceleration (\ref{eq-chi_zeta-impl}), while exact, is not given explicitly as~$\chi(\zeta)$. It is possible, however, to derive asymptotic expressions for the jet acceleration profile in the small-distance ($\zeta \ll 1$, $\chi\ll 1$) and large-distance ($\zeta\gg 1$, $\chi\to 1$) limits.
Setting $\chi \ll 1$ and taking the Taylor expansion of the $(1-\chi)^n$ terms up to third order, equation (\ref{eq-chi_zeta-impl}) is simplified to $\chi^3=3\zeta$, which can be recast in dimensional quantities as 
\beq 
u^3=\frac{6\epsilon u_\infty}{l_{\rm min}}\, z, \quad {\rm for} \quad u\ll u_\infty \, .
\label{eq-u_z-small-z}
\eeq 
This is the same as the $u\propto z^{1/3}$ solution in D02 and is independent of the $k$ parameter. {\it Therefore, below the saturation radius, the acceleration profile of the jet is universal, i.e., does not depend on $a>1$.} Physically, this is becuase this early dissipation is dominated by the annihilation of the smallest stripes in all cases. 

At large scales, $\zeta\gg 1$, we have $\chi \to 1$ and equation~(\ref{eq-chi_zeta-impl}) simplifies to $(1-\chi)^{1-k}/(k-1)\simeq\zeta$, which translates into 
\beq 
u(\zeta)=u_\infty, \{1-[(k-1)\zeta]^{1/(1-k)}\} \, , \quad u\to u_\infty \, .
\label{eq-u_z-large-z}
\eeq 

We now proceed with discussing examples of our solution for the jet acceleration, obtained by numerical inversion of equation~(\ref{eq-chi_zeta-impl}). Fig.~\ref{fig-gamma} shows the jet bulk 4-velocity $\chi = u/u_\infty$ and magnetization~$\sigma$ as functions of distance $\zeta$ for two cases: a jet with stripes of some unique width $l_{\rm min}$ (hence $a=\infty$), and a jet with a stripe-width distribution $\mathcal{P}(l)\propto l^{-5/3}$ for $l\ge l_{\rm min}$. As expected from the above analysis, the two solutions are identical at small distances ($\zeta\ll 1$). In both cases, the 4-velocities $\chi(\zeta)$ follow the $\zeta^{1/3}$ scaling (\ref{eq-u_z-small-z}) for $\zeta\ll 1$ (shown by the dotted line in Fig.~\ref{fig-gamma}); and they both reach the same asymptotic value $\chi\to 1$ at large distances. The acceleration is, however, more gradual at scales $\zeta\sim 1$ for the case of stripes with a broad range of widths. The reason for this is that the wider stripes dissipate at larger distances, making the acceleration process more gradual.   

Fig.~\ref{fig-gamma} also shows the jet magnetization $\sigma=(1-\chi)/\chi$ as a function of distance along the jet for the same two cases ($a=\infty$ and $a=5/3$). At small distances, $\zeta < 1$, the jet is Poynting-flux dominated ($\sigma\gg 1$). Gradual dissipation drives its bulk acceleration at the expense of the Poynting flux. For $\zeta\ll 1$, both example curves are characterized by $\sigma \sim 1/\chi\propto \zeta^{-1/3}$. In the case of a single stripe width $l_{\rm min}$, once the jet reaches $\zeta\gtrsim 1$, the magnetization drops steeply with distance $\sigma \sim (1-\chi)\propto \zeta^{-2}$. In contrast, for an extended distribution of stripe widths, $\mathcal{P}(l)\propto l^{-a}$, there is a more gradual decrease of the magnetization at large scales: $\sigma(\zeta \gg 1) \sim (1-\chi)\propto \zeta^{2(1-a)/(1+a)}$. For the plotted example, $a=5/3$, we have $\sigma \propto \zeta^{-1/2}$. Since, for this specific example, the dependence of $\sigma$ on distance is very similar in the small- and large-scale limits ($\sigma \propto \zeta^{-1/3}$ versus $\sigma \propto \zeta^{-1/2}$), the curve shows a rather gentle break at $\zeta\sim 1$.


\begin{figure}
\begin{center}
\includegraphics[width=0.5\textwidth]{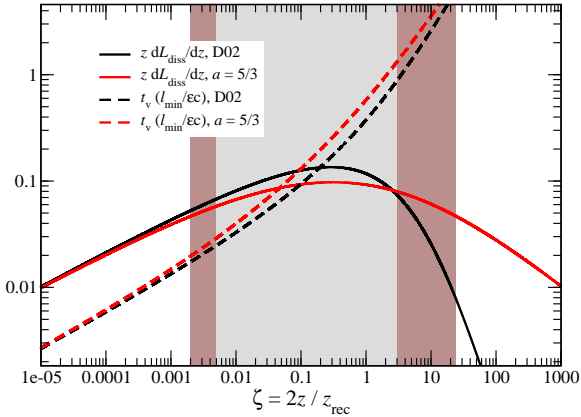}
\end{center}
\caption{Dissipation profile $z {\rm d}L_{\rm diss}/{\rm d}z$ (arbitrary units, {solid lines}) and variability time-scale $t_{\rm v}\equiv z/2\Gamma^2 c$ (in units of $l_{\rm min}/\epsilon c$, dashed lines) as functions of distance along the jet. The black (red) curves correspond to cases of a stripe-width distribution with power-law index $a\to \infty$ ($a=5/3$). The dissipation shows a broad plateau over $\sim 4$ orders of magnitude in distance. { The plateau region is shaded gray for the single stripe width case ($a\to \infty$) and brown for the $a=5/3$ case. At distance $z=z_{\rm peak}=z_{\rm rec}/6$ ($\zeta_{\rm peak} = 1/3$), the corresponding variability time-scale is simply $t_{\rm v}\simeq l_{\rm min}/12\epsilon c\sim l_{\rm min}/c$,  independent of~$\Gamma$; that is, for $\epsilon\simeq 0.1$, variability from the peak dissipation region directly tracks the dynamo timescales as they appear in the inner part of the disk.} 
\label{fig-var}
}
\end{figure}

\begin{figure}
\begin{center}
\includegraphics[width=0.5\textwidth]{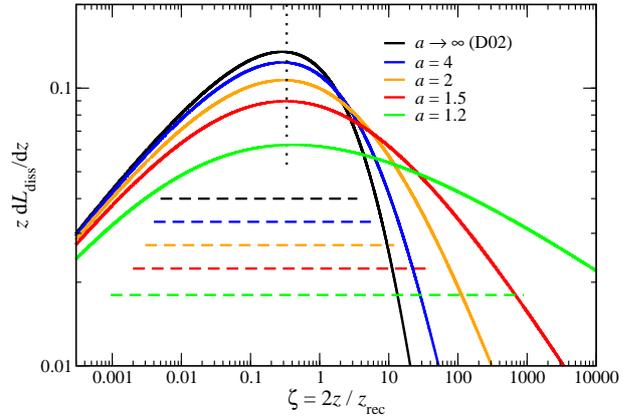}
\end{center}
\caption{Dissipation profile $z{\rm d}L_{\rm diss}/{\rm d}z$ (arbitrary units) as a function of distance along the jet. The various curves correspond to different values of the index $a$ of the stripe-width profile (see insert caption). Independently of~$a$, the dissipation peaks at $\zeta \sim 1/3$, marked with the vertical dotted line. The dissipation shows broad plateaus over $\sim 3-6$ orders of magnitude in distance. The horizontal dashed lines mark the extent of the plateaus. 
\label{fig-alpha}
}
\end{figure}


\subsection{Dissipation profile: general properties and radiative signatures}
\label{subsec-model-dissipation}

Given the jet acceleration profile, we proceed with the calculation of the dissipation rate in the jet as a function of distance. The total power dissipated per unit logarithmic interval at distance $z$ is
$z {\rm d}L_{\rm diss}/{\rm d}z = z\dot{M}c^2 {\rm d}\Gamma/{\rm d}z = z(L/\Gamma_\infty) ({\rm d}\Gamma/{\rm d}z)$. In terms of the variables $\chi$ and $\zeta$, the dissipation profile is therefore given by
\beq 
z \frac{{\rm d}L_{\rm diss}}{{\rm d}z}= L\zeta\frac{{\rm d}\chi}{{\rm d}\zeta}=L
\frac{(1-\chi)^k}{\chi^2}\zeta .
\label{eq-dissipation-profile}
\eeq 
where we used equation~(\ref{eq-chi_zeta}) and the approximation $u\simeq \Gamma$ which is valid in the ultrarelativistic regime.

For $\zeta\ll 1$ (equivalently, $\chi\ll 1$) the dissipation rate scales as $z {\rm d}L_{\rm diss}/{\rm d}z=L \zeta^{1/3}/3^{2/3}\propto z^{1/3}$, i.e., rises slowly with distance. For $\zeta \gg 1$ ($\chi\to 1$) the asymptotic solution (\ref{eq-u_z-large-z}) gives: 
$z {\rm d}L_{\rm diss}/{{\rm d}z}= L[(1+a)/(2a-2)]^{[(1-3a)/(1+a)]}\zeta^{[2(1-a)/(1+a)]} \propto z^{[2(1-a)/(1+a)]}$, i.e., a decline with distance for $a>1$, with the decline being steeper for larger $a$ values. For example, for $a= 5/3$, the decline is very gradual: $z {\rm d}L_{\rm diss}/{\rm d}z\propto z^{-1/2}$. In the limit of $a\to \infty$, we are in the case considered by D02 which gives a much steeper decline for the dissipation rate, $z {\rm d}L_{\rm diss}/{\rm d}z\propto z^{-2}$.
Interestingly, one can show that at large distances, $\zeta\gg 1$, the dissipation rate ${\rm d} L_{\rm diss}/{\rm d} \ln z\simeq L(1-\chi)^k\zeta$ in general follows the same behavior as $\sigma\simeq (1-\chi)$. 
To see this more clearly, consider the ratio of the two quantities. Using equations (\ref{eq-sigma-u}), (\ref{eq-chi_zeta}), and~(\ref{eq-dissipation-profile}), we get 
$\sigma^{-1} {\rm d}L_{\rm diss}/{\rm d} \ln z = L(1-\chi)^k\zeta/\chi$, 
which in the $\zeta\rightarrow \infty$, $\chi\rightarrow 1$ limit becomes, using equation~(\ref{eq-u_z-large-z}), $\sigma^{-1} {\rm d} L_{\rm diss}/ {\rm d} \ln z\simeq L(1-\chi)^{k-1}\zeta = L/(k-1)$, i.e., it is independent of distance~$\zeta$.
Figs.~\ref{fig-var} and~\ref{fig-alpha} show the dissipation profile in the jet for various values of~$a$. Fig.~\ref{fig-alpha} demonstrates that, independently of~$a$, the dissipation peaks at $\zeta\simeq 1/3$ which motivates us to define the "peak" dissipation distance 
\beq 
z_{\rm peak}\equiv \frac{z_{\rm rec}}{6}=\frac{\Gamma_{\infty}^2l_{\rm min}}{6\epsilon}.
\eeq

The peak in the dissipation rate is more of a plateau than a well-defined maximum. To quantify the width of the plateau, we measure the range of distances for which the dissipation rate remains within 50\% of its peak value. For $a\to \infty$, the plateau extends over the interval $5\times 10^{-3}\lesssim \zeta\lesssim 3$, while for higher values of $a$ the plateau is even broader, e.g., $2\times 10^{-3}\lesssim \zeta\lesssim 20$ for $a=5/3$ (see the shaded regions of Fig.~\ref{fig-var}). The dissipation profile and the extent of the plateau for different values of the $a$ parameter are shown in Fig.~\ref{fig-alpha}. 
{\it As is clear from this Figure, rather independently of the $a$ parameter, the model predicts an essentially flat dissipation profile that extends over about $4-5$ orders of magnitude in distance.} 
In particular, dissipation remains substantial at distances 2 orders of magnitude smaller or larger  than~$z_{\rm peak}$. Over this range of distances, the jet accelerates from $\sim \Gamma_\infty/6$ to its terminal speed and, correspondingly, the jet magnetization drops from $\sigma\simeq 5$ to $\sim 0.1$ (see also Fig.~\ref{fig-gamma}).


The very broad dissipation profile also has profound implications for the jet emission time variability (see Fig.~\ref{fig-var}). For an observer with the line of sight along the jet axis, a dissipation event that sets in at distance $z$ and proceeds until distance $\sim 2z$ produces a flare of an observed variability time-scale $t_{\rm v}\simeq z/2\Gamma^2c$. 
For $z=z_{\rm peak}=\Gamma_\infty^2l_{\rm min}/6\epsilon$, the corresponding variability time-scale is simply $t_{\rm v}\simeq l_{\rm min}/12\epsilon c\sim l_{\rm min}/c$,  independent of~$\Gamma$; that is, for $\epsilon\simeq 0.1$, variability from the peak dissipation region directly tracks the dynamo time-scales as they appear in the inner part of the disk! Dissipation at the outer plateau results in variability time-scales about two orders of magnitude longer. Overall the model predicts powerful flares that extend in time over $0.1\lesssim t_{\rm v}/[l_{\rm min}/\epsilon c]\lesssim 10$ (Fig.~\ref{fig-var}). 
Adopting the reference values $l_{\rm min}=100R_g$, $\epsilon =0.1$, we see that this range translates to jet flaring time-scales in a broad range of $10^2\lesssim R_g/c\lesssim 10^4$. In the next Section we explore what these estimates imply for astrophysical jets.


\section{Astrophysical Implications}
\label{sec-astro}

So far, we explored the general properties of the striped jet model. We now proceed to apply the model to two specific classes of sources: blazars and GRBs. The model developed here is, in principle, also relevant for other systems that contain relativistic jets driven by accreting black holes, such as microquasars. In practice, however, such jets are likely to be moderately relativistic $\Gamma\sim 2$ (see, e.g., Remillard \& McClintock 2006), limiting the applicability of our model which explicitly considers ultrarelativisitc, $\Gamma\gg 1$, flows.


\subsection{Blazars}
\label{subsec-astro-blazars}

In this subsection we apply our model to jets from supermassive black holes in~AGN. We consider a black hole of mass $M\sim 10^8 - 10^9M_\odot$ surrounded by an accretion disk. The jet is launched at a $\sim R_g$ scale and undergoes magneto-centrifugal acceleration and collimation from interaction with the surrounding accretion disk wind. We assume that the  magnetic field polarity through the inner disk and the black hole changes on timescales $t\gtrsim l_{\rm min}/c\sim 1000R_g/c$, as discussed in Section~\ref{sec-picture}. Therefore, the effects of the polarity reversals are not important for the jet dynamics at the compact launching and collimation scales. Ideal MHD acceleration is expected to dominate at these distances. The focus of this work is on the dissipative acceleration that takes place at larger scales. We thus set the inner boundary for our calculation at a distance of $z_{\rm o}=100R_g$ where the jet is assumed to have an initial Lorentz factor $\Gamma_{\rm o}=3$. The choice of $z_{\rm o}$ and $\Gamma_{\rm o}$ is rather ad hoc. We have checked, however, that our results do not depend sensitively on this choice as long as $z_{\rm o}\lesssim 3\times 10^3R_g$ and $\Gamma_{\rm o}\lesssim 5$. The asymptotic jet 4-velocity is chosen to be  $u_\infty\simeq \Gamma_\infty =30$. For the example presented here, we adopt a power-law index for the stripe width~$a=5/3$.

\begin{figure}
\begin{center}
\includegraphics[width=0.5\textwidth]{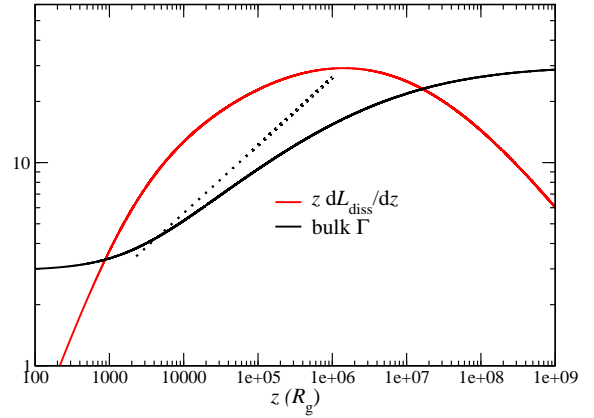}
\end{center}
\caption{Example of an AGN jet solution with asymptotic Lorentz factor $\Gamma_\infty=30$ {and $a=5/3$}. The black (red) curve shows the Lorentz factor (dissipation profile). The dotted line shows the D02 analytic solution, applicable scales $z_{\rm o}\ll z\ll z_{\rm rec}$ (eq.~\ref{eq-u_z-small-z}).
\label{fig-blazar}
}
\end{figure}

Fig.~\ref{fig-blazar} shows the acceleration and dissipation profiles of the jet. For our reference parameters the reconnection scale is $z_{\rm rec}=\Gamma_\infty^2l_{\rm min}/\epsilon \simeq 10^{7}R_g$  (for $\epsilon=0.1$), giving a peak in the dissipation rate at distance $z_{\rm peak}= z_{\rm rec}/6\simeq 10^6R_g$. The dissipation profile is essentially flat from $\sim 10^4R_g$ to about~$10^8R_g$. For an $M\sim 10^8M_\odot$ black hole, this corresponds to scales $10^{17}-10^{21}$~cm (or $0.03-300$~pc). In general, keeping the dependence of dissipation on $\Gamma_\infty$, 
\beq 
z_{\rm peak}\simeq 10^6\left(\frac{\Gamma_\infty}{30}\right)^2\frac{l_{\rm min}}{1000R_g}R_g.
\eeq 
One arrives at the same range of scales if one considers a larger black hole mass $M\sim 10^9M_\odot$ but smaller $l_{\rm min}=100R_g$. Slower jets give smaller peak dissipation scale. 
A number of blazar observables may be understood in the context of this model. 

Homan et al. (2015) published a systematic study on the acceleration properties of blazar jets from the MOJAVE Very Long Baseline Array program. They found that about three-quarters of the jets in their sample show significant changes in the Lorentz factors, namely, a broad trend of jet features increasing their speed due to an acceleration of the jet flow out to distances of the order of $100$~pc. They estimate that the jet Lorentz factor changes by a factor of a few over these length scales. This is in agreement with our model, which is characterized by a prolonged acceleration of the jet. In the example shown in Fig.~\ref{fig-blazar}, the jet accelerates from $\Gamma=\Gamma_\infty/2$ to $0.9 \Gamma_\infty$ over a distance range of about $10^6-10^8 R_g$. Note that the geometric mean of this range ($10^7R_g$) corresponds to a scale of $\sim$150~pc for a supermassive black hole mass of $M\sim 3\times 10^8M_{\odot}$. 

Blazar jets are variable over a broad range of time-scales. Major flares last from a few hours to several months (see, e.g., Krawczynski et al. 2004). Occasionally, powerful flares of duration as short as $\sim$10 minutes have been observed in the X-ray and $\gamma$-ray bands (Cui 2004; Albert et al. 2007; Aharonian et al. 2007; see also Begelman, Fabian \& Rees 2008). In the context of the striped jet model, variability time-scales range from $\sim l_{\rm min}/c\sim 100R_g/c\sim 1$ day at distance $z\sim z_{\rm peak}$ to $\sim 100\,l_{\rm min}/c\sim 10^4R_g/c\sim 100$ days at the outer edge of the plateau zone. This may help us understand why major flares from a single source have such a broad range to time-scales.  Moreover, dissipation at the inner edge of the plateau zone $z\sim 10^4 R_g$ is characterized by $\sigma \sim 5\gg 1$. There, magnetic reconnection may drive relativistic bulk fluid flows in the jet comoving frame (Lyubarsky 2005) giving rise to even faster evolving flares, i.e. the "jets-in-a-jet" model, characterized by observed variability time-scales even shorter than~$R_g/c$ (Giannios, Uzdensky \& Begelman 2009; 2010; Nalewajko et al. 2011; Giannios~2013).

Optical polarimetry monitoring indicates that major $\gamma$-ray flares may be commonly accompanied by major polarization angle swings (Blinov et al.~2018). Such swings may be easier to understand in a model where (i) the jet contains stripes of reversing toroidal magnetic field, and (ii) dissipation in the current sheets that separate the regions of polarity reversals is responsible for the blazar activity. 

A conical jet with flat dissipation profile can account for the flat-to-inverted radio emission spectrum observed in many jetted sources~(BK79). The BK79 model considers a conical jet coasting with constant velocity and with its magnetic field decreasing with distance as $B\propto 1/z$. The jet contains relativistic electrons, in equipartition with the magnetic field over a broad range of lengthscales. The turnover frequency due to synchrotron self-absorption scales as $\nu_t\propto 1/z$ resulting in a jet that becomes transparent at larger scale for lower frequency. The resulting synchrotron spectrum is flat.  Critical for the BK79 model to work is the re-acceleration of the radiating electrons over a very broad range of distances. Absent of this re-acceleration, adiabatic expansion would result in energy loss and little extended emission in the jet. The striped jet model discussed here naturally explains how particles keep being accelerated over a very broad range of distances at an almost constant rate.  The striped jet model also predicts approximately the comoving magnetic field dependence on distance that is considered in the BK79 model. For $\zeta\gtrsim 1$, $\Gamma\simeq \Gamma_\infty$ and $B\propto \sqrt{\sigma}/z\propto z^{-2a/(1+a)}$. For instance for $a=5/3$ the magnetic field scales as $B\propto z^{-5/4}$, i.e., fairly close to the BK79 $B\propto 1/z$ scaling. 
Furthermore, magnetic reconnection results in approximate equipartition between radiating particles and magnetic field (e.g., Sironi, Petropoulou \& Giannios 2015; Sironi, Giannios \& Petropoulou 2016; Werner \& Uzdensky~2017; Werner et al.~2018). Therefore, although the detailed predictions of the model still need to be explored, it provides a promising explanation for the typically observed  flat-to-inverted long-wavelength emission from AGN jets.


\subsection{Gamma-Ray Bursts}
\label{subsec-astro-GRBs}

Here we explore the implications of the model for GRB jets, for both classes of these fascinating  objects: the long-duration GRBs, believed to be powered by the gravitational collapse of massive stars (the collapsar scenario; see, e.g., Woosley 1993), and the short-duration GRBs, believed to originate from mergers of two neutron stars (e.g., Eichler et al. 1989, see Baiotti \& Rezzolla 2017 for recent review). 
We assume that the GRB central engine that naturally forms in both cases is an accreting black hole of several solar masses (such that $R_g=10^6$cm) surrounded by an accretion disk.  
We take the minimum stripe width of $l_{\rm min}=1000R_g\simeq 10^9$cm, and asymptotic velocity
$u_\infty\simeq \Gamma_\infty \simeq 300$ as normally inferred for~GRBs. The jet is launched from the black hole and undergoes ideal MHD acceleration and collimation while it propagates through the collapsing star (for long-duration  GRBs) or the merger ejecta (for short GRBs). The jet is assumed to have a Lorentz factor $\Gamma_{\rm o}=20$ at the breakout distance of $z_{\rm o}=3\times 10^{10}$cm, comparable to the size of the progenitor star. After breakout, the jet is no longer confined by the surrounding gas and can therefore expand freely, turning conical. The scale $z_{\rm o}$ serves as the injection point for the jet. Our results do not depend sensitively on the choice of these parameters as long as $z_{\rm o}\lesssim 3\times 10^{11}$cm and $\Gamma_{\rm o}\lesssim 50$. From these fiducial parameters, the reconnection scale is $z_{\rm rec}\simeq 10^{15}$cm, while the peak in the dissipation rate is at $z_{\rm peak}= z_{\rm rec}/6\sim 10^{14}$~cm. We set the power-law index for the stripe width distribution to~$a=5/3$. The dissipation profile is then found to be essentially flat from $\sim 10^{12}$~cm to~$10^{16}$~cm. Acceleration proceeds out to scales of $\sim 10^{16}$~cm, just before the afterglow stages set in. Keeping the dependence of dissipation on $\Gamma_\infty$, for $\epsilon=0.1$, we find
\beq 
z_{\rm peak}\simeq 10^{14}\,{\rm cm}\left(\frac{\Gamma_\infty}{300}\right)^2\frac{l_{\rm min}}{1000R_g}.
\eeq 

\begin{figure}
\begin{center}
\includegraphics[width=0.5\textwidth]{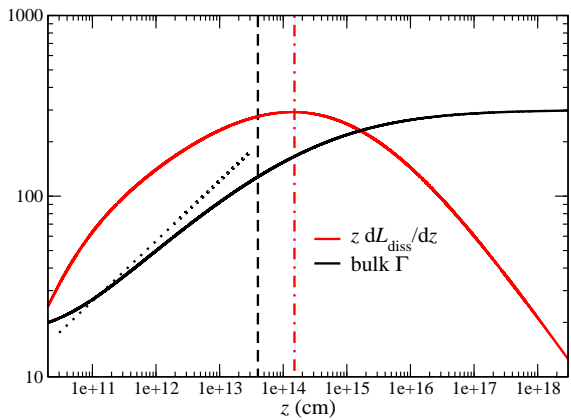}
\end{center}
\caption{Example of a GRB jet with asymptotic Lorentz factor $\Gamma_\infty=300$ { and $a=5/3$}. 
The black (red) curve shows the Lorentz factor (dissipation profile). The vertical dashed line marks the location of the Thomson photosphere for jet luminosity $L=10^{53}$~erg s$^{-1}$. The dotted line shows the D02 analytic solution, applicable scales $z_{\rm o}\ll z\ll z_{\rm rec}$ (eq.~\ref{eq-u_z-small-z}). The vertical red, dash-dotted line shows the peak dissipation distance  $z_{\rm peak}=1.5\times10^8R_{\rm g}=1.5\times 10^{14}$~cm.
\label{fig-GRB}
}
\end{figure}

Unlike blazar jets, GRB jets are opaque to electron scattering at compact scales. The Thomson optical depth $\tau_T$ along the jet propagation direction can be estimated by considering the electron rest-frame number density $n$ and the (comoving) expansion length of the jet $z/\Gamma$, giving $\tau_{\rm T} \simeq n\sigma_{\rm T} z/\Gamma$, where $\sigma_T$ is the Thomson scattering cross section. 
This expression can be solved for the $\tau_{\rm T}=1$ photospheric radius as $z_{\rm ph}=4\times 10^{13} L_{53}\Gamma_{\infty,300}^{-1}\Gamma_{100}^{-2}$~cm, where the jet luminosity is $L=10^{53}L_{53}\, {\rm erg}\ {\rm s}^{-1}$ (e.g., Giannios 2012). For these reference values, $z_{\rm ph}\simeq 4\times 10^{13}$cm as marked with a dotted line in Fig.~\ref{fig-GRB}, i.e., somewhat below the peak dissipation distance.


The model shown in Fig.~\ref{fig-GRB} is thus characterized by a powerful photospheric component (approximately 40\% of the energy is dissipated below and around the photosphere). Such a dissipative photosphere has a non-thermal appearance that may account for much of the GRB prompt emission (Thompson 1994; M{\'e}sz{\'a}ros \& Rees 2000; Pe'er, M{\'e}sz{\'a}ros, \& Rees 2006; Giannios 2006, 2008; Beloborodov 2010; McKinney \& Uzdensky 2012). Dissipation, however, continues also outside the photosphere, throughout the optically thin region, out to $10^{16}$cm scales. Non-thermal particles can produce synchrotron emission component that contributes to the GRB (Lyutikov \& Blandford 2003; Giannios \& Spruit 2005; Lyutikov 2006; Zhang \& Yan 2011; Beniamini et al. 2018). Interestingly, the prompt emission spectra of a significant fraction of bursts are best described by two components: a narrow one, peaked at $\sim~100$~keV, and a broader component (see, e.g., Guiriec et al. 2017). The first component has been interpreted to originate from the Thomson photosphere of the jet (e.g., Ryde 2005), while the second one may be attributed to synchrotron emission from larger scales in the jet.

GRB light-curves are observed to be strongly variable. They are made up of pulses that may last for up to a second and contain much faster (ms-long) sub-structures (see, e.g., Fenimore et al. 1995). Interestingly, the characteristic variability time-scale associated with $z\sim z_{\rm peak}$ is $t_{\rm v}\simeq l_{\rm min}/c\sim 10^3R_g/c\sim 0.03$~s. This time-scale may be related to the typical duration of a GRB pulse, although the latter is typically an order of magnitude longer. On the other hand, dissipation around the Thomson photosphere of the jet can lead to variability on $1-10$~ms time-scales, comparable to the shortest observed variability time-scales in the bursts. Dissipation at the outer edge of the plateau region leads to $t_{\rm v}\sim$~10~s; possibly related to the overall duration of operation of the central engine. 

A caveat for the applicability of our model to GRBs may come from the limited size of the accretion disk in these sources, resulting in a limited dynamic range of field-reversal time-scales. This is in contrast to disks in AGN which are thought to be very extended. Indeed, the outer extent of the disk $R_{\rm out}$ depends sensitively on the distribution of the specific angular momentum of the GRB progenitor system and may be characterized by $R_{\rm out}/R_{\rm base}\sim 1-30$ {for collapsar disks} (e.g. Woosley 1993) and $R_{\rm out}/R_{\rm base}\sim$ a few for neutron-star mergers. In our analysis, we assume that the stripe width distribution extends to $l\gg l_{\rm min}$. 
{For a finite disk, a more realistic stripe width distribution in the jet should be truncated at some maximum~$l_{\rm max}$. Since the stripe width $l$ associated with a disk distance $R$ scales as $l(R)\propto\Omega_K^{-1}(R) \propto R^{3/2}$, we can estimate $l_{\rm max}$ as $l_{\rm max}\sim (R_{\rm out}/R_{\rm base})^{3/2}l_{\rm min}$.}
In practice, for $R_{\rm out}/R_{\rm base} \gtrsim 5$, we have $l_{\rm max}\gg l_{\rm min}$, and the error due to not truncating the $l$ distribution at a finite $l_{\rm max}$ value is rather small. 

On the other hand, a rather narrow disk, with say $R_{\rm out}/R_{\rm base} \sim 1- 2$ may be better described by the D02 single stripe width model. Such model can be particularly relevant for the compact disk/torus expected to form in binary neutron star mergers and to give rise to short-duration GRBs. The D02 model predicts a sharper drop of the dissipation rate with distance above $z_{\rm peak}$. In a model with identical parameters to that discussed in the beginning of this section, but with $a=\infty$, the dissipation plateau extends over distances ranging from $\sim 10^{12}$~cm to~$10^{15}$~cm, i.e., the outer edge by a factor $\sim$10 shorter than in the $a=5/3$ case. Correspondingly, all else been equal, the short GRB jets will be characterized by dissipation at rather more compact scales and a stronger photospheric component\footnote{The mass of the black hole in the merger scenario is likely to be a factor of a few smaller than that of the collapsar, resulting in a smaller gravitational radius for the former, making the dissipation scale for short GRBs even more compact in comparison to that of long bursts.}. 
Photospheric emission spectra are particularly hard below the peak of the $\nu f_\nu$ spectrum. Interestingly, short-duration GRBs are spectrally harder than long GRBs (Kouveliotou et al. 1993).    


\section{Discussion}
\label{sec-disc}

Relativistic jets are likely to be driven by strong magnetic fields threading the black hole and/or the inner accretion disc (Blandford \& Znajek 1977; Blandford \& Payne 1982). However, the origin of the magnetic fields remains debated. Large-scale magnetic fields may be dragged inward by the accreting material  over several orders of magnitude in distance towards the black hole  (e.g., Livio, Ogilvie, \& Pringle 1999; Spruit \& Uzdensky 2005;  Rothstein \& Lovelace 2008; Beckwith, Hawley, \& Krolik 2009; Guilet \& Ogilvie 2012; see, however, Lubow, Papaloizou \& Pringle 1994; Heyvaerts, Priest \& Bardou 1996; Guan \&  Gammie 2009). As a result, the fields are compressed, becoming sufficiently strong to account for the observed jet power (e.g., Tchekhovskoy, Narayan  \& McKinney 2011). In this picture, the polarity of the magnetic field through the black hole and, therefore, the jet cross section is fixed. The jet acceleration and emission stages are, in general, decoupled. The jet accelerates through ideal MHD processes converting part of its magnetic energy into bulk motion. The energy dissipation, particle acceleration and emission of electromagnetic radiation can take place at larger scales. The dissipation may result from either MHD instabilities (Eichler 1993; Moll et al. 2008; McKinney \& Blandford 2009; McKinney \& Uzdensky 2012; Bromberg \& Tchekhovskoy 2016; Barniol-Duran, Tchekhovskoy \& Giannios 2017) or, if the acceleration process is very efficient and the jet flow becomes kinetic-energy dominated, through shock waves (Rees \& Meszaros 1994).

In this work, we consider the case where the magnetic fields responsible for launching jets are generated and amplified locally in the disc through the turbulent MRI-driven dynamo (Davis et al. 2010, O'Neill et al. 2011, Simon et al. 2012). This implies that (i) the magnetic field polarity through the disc, and, consequently, the black hole, is changing with time and (ii) the changes reflect the MRI growth timescales in the accretion disc. The result is a striped jet. {\it An important property of such a  jet is that dissipation in the jet and bulk acceleration track each other (D02).} Most of the gravitational energy is released in the inner disc where the characteristic MRI dynamo time-scale is $\tau_{\rm min}\sim (100-1000)R_g/c$. 
{As a result, the magnetic field at the jet base reverses polarity on time-scales $\tau\gtrsim \tau_{\rm min}$; correspondingly, the jet contains stripes of a characteristic width $l_{\rm min}\sim c\tau_{\rm min}\sim (100-1000)R_g$, but with the stripe width distribution extending to larger scales.} As we have shown {in this article}, the distribution of the stripe widths plays a crucial role in determining the dissipation and acceleration profiles of the jet {at large distances}.   

The determination of the stripe width distribution is a challenging theoretical problem. It requires global MHD disc simulations with sufficient resolution to resolve the MRI, in combination with analytical theory. 
In astrophysical systems of interest the disk can be very extended (AGN, black-hole binaries) or more compact (GRBs). Discs can be hot and geometrically thick, characterized by rapid accretion of matter, or cooler and geometrically thin. The distributions of length-scales introduced by the MRI dynamo may be different in these different systems.

Some answers to these issues can come from observations. Not all accreting black holes show evidence of relativistic jets. Radio-quiet AGN and black-hole binaries in soft spectral states are prominent examples of jet-less black hole systems (Remillard \& McClintock 2006). A natural question that arises, therefore, is: why is the presence of an accretion disc not a sufficient condition for a powerful jet? After all, the MRI is believed to operate in all black-hole accreting systems. We suspect that the answer is that either (i) jet-less systems have black holes of slow spin, resulting in inefficient Blandford-Znajek process or (ii) the inner part of the disc in these systems is geometrically thin and advection of the magnetic field in thin discs is inefficient. Further studies of the radio-loud and radio-quiet AGN, and state transitions of X-ray binaries combined with black-hole spin inferences in these systems may help to clarify the role of the disc thickness and the black-hole spin in launching relativistic jets.


\section{Conclusions}
\label{sec-conclusions}

In this work, we developed a formulation (see equations~\ref{eq-pl-int}, \ref{eq-accel-diff}) with which one can calculate the dissipation and acceleration profiles of a steady, conical {relativistic} jet that contains stripes of reversing magnetic field polarity which follow a distribution of widths $\mathcal{P}(l)$. As the jet expands, progressively wider stripes dissipate, converting their magnetic energy into bulk motion and energetic particles. The particles then radiate producing the observed electromagnetic radiation.
As we showed in this study, for a stripe-width distribution dominated by the scale $l_{\rm min}$, the dissipation rate in the jet peaks at  $z_{\rm peak}\simeq \Gamma_{\infty}^2l_{\rm min}/6\epsilon$ but remains substantial over a broad range of distances (the plateau) that extends for two orders of magnitude to both shorter and longer scales. For jets moving close to the observer's line of sight, we estimate that the characteristic variability time-scale corresponding to the peak dissipation distance is $t_{\rm v}\simeq z_{\rm peak}/2\Gamma_{\infty}^2c= l_{\rm min}/12 \epsilon c$. The variability timescale extends to a factor of 100 longer timescales for dissipation at the outer edge of the plateau region. 

When applied to blazars, assuming a jet that is accelerated to a bulk $\Gamma_\infty=30$, our model predicts that the dissipation rate peaks at about a few parsec away from the supermassive black hole. The actual `blazar zone', however, is a very extended region starting from $\sim 0.03$~pc and extending out to hundreds parsecs. The corresponding variability time-scales of flares vary from $\sim$1 day to several months. Faster evolving flares are possible in the inner part of the blazar zone where the jet is magnetically dominated and the jets-in-a-jet {scenario} is applicable. The model also predicts that dissipation of magnetic energy drives the bulk acceleration of the jet. Blazar jets are, therefore, expected to accelerate out to scales of several hundreds~pc.

For GRB-relevant parameters, a jet with asymptotic Lorentz factor of $\Gamma_\infty=300$ is characterized by $z_{\rm peak}\sim 10^{14}$~cm. Approximately half of the magnetic energy is dissipated under optically thick conditions, contributing to the photospheric emission component, while the rest of the dissipation may manifest itself as optically thin emission. Mergers of neutron stars powering short-duration GRBs may launch jets with somewhat shorter stripe widths, resulting in a more compact dissipation region and, correspondingly, a stronger photospheric component.  

This study provides a quantitative connection between the accretion disk variability on the one hand and the jet dynamics and dissipation profile on the other. This model can facilitate the bridging of physical processes that take place at the inner disk (a few $R_{\rm g}$) to observable signatures at vastly larger ($\sim 10^6R_{\rm g}-10^{10}R_{\rm g}$) scales where the jets radiate. Future progress may be possible from both directions.  Studies of the accretion physics can provide us with an accurate distribution of widths of the stripes $\mathcal{P}(l)$ and therefore more accurate models for the jet dissipation, acceleration, and emission profiles. On the other hand, observations resolving the distribution of power as function of distance from the black hole may help us to make inferences of how accretion disc operates close to the black hole.


\section*{Acknowledgements}

DG acknowledges support from NASA grants NNX16AB32G and NNX17AG21G.
DAU acknowledges support from DOE grant DE-SC0008409, NASA grants NNX16AB28G and NNX17AK57G, and NSF grant AST-1411879.








\bsp	
\label{lastpage}
\end{document}